# Optical Spectra of Open-Shell and Closed-Shell Graphene-Based Molecules

Mikhail F. Budyka[X], Elena F. Sheka*[§] and Nadezhda A. Popova[§]

E-mail: sheka@icp.ac.ru
*To whom correspondence should be addressed
[X]Institute of Problems of Chemical Physics of RAS, Chernogolovka, Russia
[§]Peoples' Friendship University of Russia, Moscow, Russia

**Abstract**
We have computationally investigated absorption spectra of a specifically configured set of graphene-based molecules involving (1) a $sp^2$ bare graphene sheet; (2) framed graphene sheets containing different chemical addends terminating dangling bonds of edge atoms but keeping $sp^2$ configured basal plane; and (3) 'bulk' $sp^3$ graphene sheets resulted from the chemical modification occurred at not only the bare sheet circumference but at its basal plane as well. Framed molecules, open-shell by nature, present different kinds of reduced graphene oxides and present the main building blocks of graphene quantum dots. Closed-shell 'bulk' molecules present models of nanosize graphene oxide. UHF ground states and ZINDO/S excited states of the molecules were analyzed. The UHF-ZINDO/S combination is well coherent in the case of 'bulk' molecules for which UHF and RHF ground state results are identical. In the case of framed molecules, the incoherence of the UHF and close-shell ZINDO/S approaches is revealed in a considerable decreasing of the HOMO-LUMO energy gap (($G_{RHF}/G_{UHF} \ll 1$), which provides unreal drastic red shift of absorption spectra of open-shell molecules into IR region. The conclusion is justified by a comparative analysis of calculated and experimental data on absorption and fluorescence spectra of graphene quantum dots and graphene oxides.

**Keywords:** graphene oxide, reduced graphene oxide, excited states, open-shell and closed-shell molecules; unrestricted Hartree-Fock and ZINDO/S approaches, absorption and fluorescence spectra

## 1. Introduction

Luminescence stability, nanosecond lifetime, biocompatibility, low toxicity, and high water solubility make graphene quantum dots (GQDs) excellent applicants for opto-electronics [1-3] and biophysics [4-7], particularly, high contrast bioimaging and biosensing. The latter stimulated the growth of interest in GQD, in general, and their preparation, in particular. Meeting this demand, a lot of synthetic methods appeared to produce GQDs, both 'top-down' and 'bottom-up' that are described in a number of reviews [2, 8, 9]. Analysis of structure and chemical compositions show that in all the cases GQDs are a few layer stacks of reduced graphene oxide (rGO) sheets of 1-10 *nm* in lateral size. rGO stacks are highly varied differing by both the number of layers and linear dimension of the sheets as well by chemical composition and shape of the latter. Thus, the variety of samples resulting from the reduction process turns out to be extremely sensitive to structural details, as well as to the amount and type of oxidation, as demonstrated by a number of studies on the vibrational properties of such systems [11-17] as well as their absorption and emission spectra [18-24]. In contrast to extended experimental studies, computational consideration of GQDs spectral properties is highly scarce. One of the main reasons of the situation is the fact that rGO chemical complexity and uncertainty in size and shape of the relevant sheets inhibit the possibility of identifying a unique model structure representative of the species. Against this background, meeting the demands on definite data information, structural-spectral investigations of shungite natural GQDs occupy a special place in the total mass of the GQDs research. Empirically determined size and chemical composition of shungite GQDs [10-13, 25, 26], their vibrational [11-13] and absorption-fluorescence [26-28] spectra provide unique possibility to adequately simulate not only the GQD ground state but its excited states as well. The next reason is related to the rGO open-shell character that raises issues concerning computational tools in use.

The current paper presents results of an extended computational experiment that, on the one hand, is based on a set of model structures, which take into account the available spectral-structural information about shungite GQDs, while, on the other hand, extends the rGO model set including parent models of closed-shell graphene oxide (GO) to reveal particular points of the excited states calculations, the most sensitive to the open-shell/closed-shell transformation. All the studied models are polyderivatives of the same parent graphene molecule with different oxygen-based groups covalently bonded either to the edge atoms only, thus matching a set of different rGOs, or also to the basal plane atoms, thus transforming rGOs to GOs. Since, in both cases, the sheets are weakly bound in the stacks, from the optical spectroscopy viewpoint, not stacks themselves but the rGO and GO fragments determine PL properties of the final products due to which we can neglect the sheet aggregation and concentrate on individual rGO and GO molecules when studying spectral properties of GQDs and GO aggregates. Absorption spectra were the main subject of the study with an accent on the validity of computational tools traditionally used, about what may be judged by comparing the calculated and experimental spectra.

## 2. Objects under study

Generally, graphene-based molecules can be divided into three groups: (i) verily graphene (pristine) molecules presenting pieces of bare flat honeycomb sheets with non saturated dangling bonds of edge atoms; (ii) framed graphene molecules that are the molecules with fully or partially saturated dangling bonds in the circumference area; and (iii) 'bulk' graphene molecules with chemical addends enveloping the whole body of the molecule carbon skeleton. The framed-and-'bulk' division reveals the unique two-zone feature of the chemical activity of pristine graphene molecules that governs the formation of any of their derivatives [29]. The (5, 5) nanographene (NGr) molecule in Fig. 1a presents a rectangle graphene sheet (55gr below) with bare edges involving five benzenoid units along both armchair and zigzag edges. Its 'chemical portrait' is shown in Fig.1b in view of the atomic chemical susceptibility (ACS) image map [30]. As seen in Fig. 1b, the map clearly reveals two zones, namely, circumference covering edge atoms and basal plane, ACS of which, expressed via the number of effectively unpaired electrons on atoms $N_{DA}$, differs more than four times in the favor of the former ones. Accordingly, the framing of pristine (bare) graphene molecules is the first stage of any chemical modification towards a complete polyderivatization. Since this reactivity area is largely spread in space, the formation of the first monoderivative does not inhibit the molecule reactivity so that the reaction will continue until the reaction ability is satisfied. This means that any chemical modification of graphene starts as saturated polyderivatization of the pristine molecule at its circumference. Consequently, involving basal plane atoms into reaction occurs usually at the second stage. The ACS image map in Fig. 1b evidences as well the open-shell character of the pristine 55gr molecule caused by the correlation of its $p_z$ odd electrons [30]. The total number of effectively unpaired electrons $N_D = \sum_A N_{DA}$ presents the measure of the correlation extent.

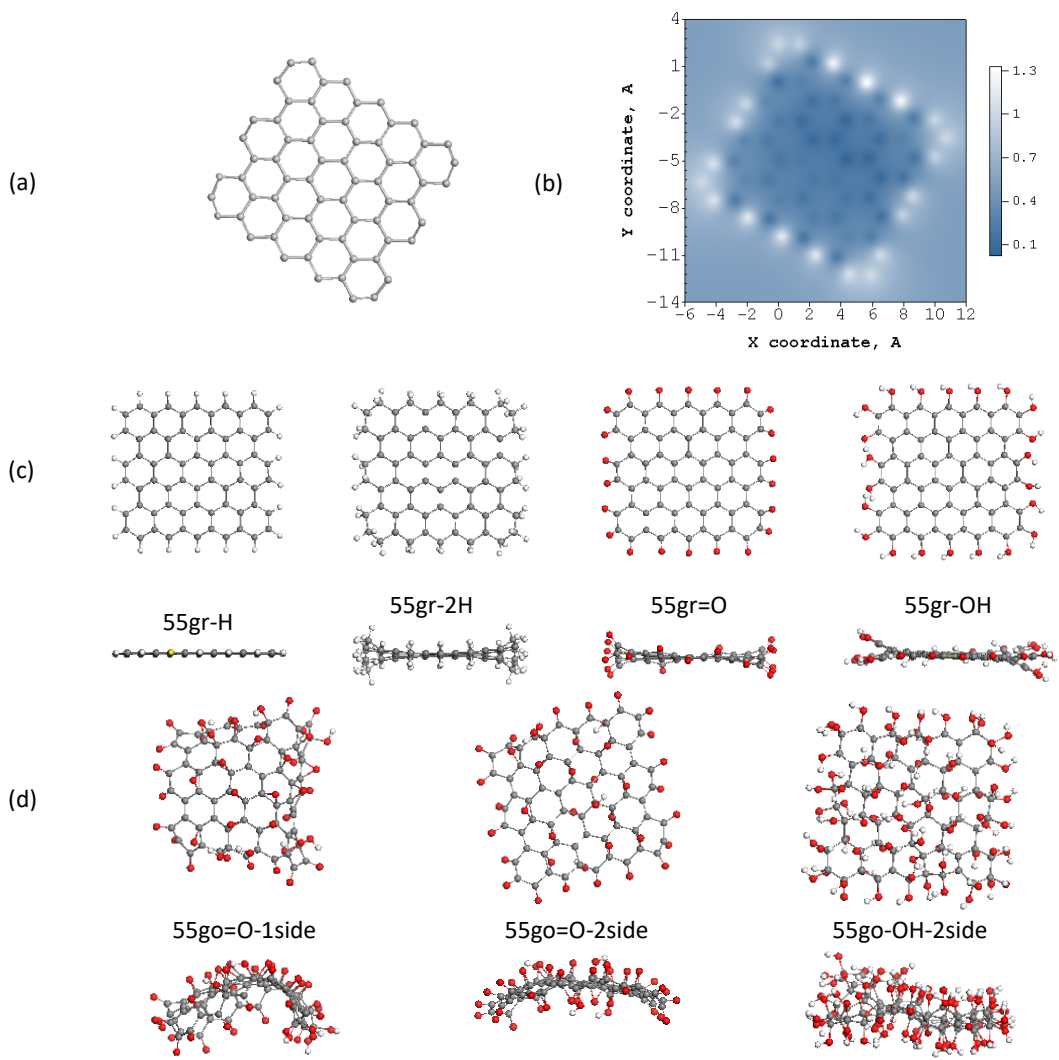

**Figure 1.** Ball-and-stick representation of the equilibrium structures of studied molecular models (top and side views). C atoms are depicted in grey, H in white and O in red. (a) Pristine (5, 5) NGr molecule; (c) (5, 5) NGr framed with (mono)(55gr-H) and (di) (55gr-2H) atomic hydrogen, atomic oxygen (55gr=O) and hydroxyls (55gr-OH); (d) Framed 55gr=O and 55gr-OH with epoxy (one- and two-side) and hydroxyl (two side) units on the basal plane, 55go=O (1side and 2side) and 55go-OH, respectively. (b) Atomic chemical susceptibility $N_{DA}$ image map over (5, 5) NGr molecule atoms in real space. Scale bars match $N_{DA}$ values. UHF AM1 calculations.

The chemical complexity of both GO and rGO, mentioned above and making impossible suggestion of a unique model structure representative of the species [8], forces to limit ourselves by a restricted number of characteristic molecular models. Presented in Figs. 1c and 1d is an example of the models subordinated to a particular choice governed by structural and chemical data obtained for shungite rGO (rGO-Sh) and summarized in [31]. As shown, rGO-Sh presents sheets of linear size from the low part of nanometer scale with the uppest bound evaluated as 3-4 *nm*. Its chemical composition is described by an average per-one-hexagon formula $C_6O_{0.1}H_{1.6-0.7}$. The framed molecules in Fig. 1c, presenting rGO models in our study, have been selected to fit the data. The linear sizes of pristine 55gr along armchair (1.12 *nm*) and zigzag (1,22 *nm*) provide obtaining framed rGO structures with linear size below 3 *nm*. 55gr-H molecule with atom-content formula $C_6H_2$ suits the best chemical composition of rGO-Sh while 55gr-2H, 55gr=O and 55gr-OH are selected to investigate how violated atom-content composition of rGOs at preserved carbon skeleton may affect final results. Additionally, framing the carbon 55gr-H skeleton by one hydrogen per each edge atom leaves the circumference valently unsaturated while addition of the second hydrogen fully inhibits its chemical activity. The inhibition is preserved when the framing involves oxygen atoms in the case of 55gr=O molecule while becoming again valently unsaturated in the case of 55gr-OH. Therefore, a set of four models in Fig. 1c allows revealing three effects on the absorption spectra under the question: (i) valence saturation of the edge atoms; (ii) change in the chemical compositions in the case of both valently saturated (2H and O) and unsaturated ( H and OH); (iii) the electron correlation in the molecules expressed by $N_D$ values.

'Bulk' molecules in Fig. 1d present three (5, 5) GOs formed in the course of the further monooxidant treatment of 55gr=O and 55gr-OH rGO molecules transferring them into 55go=O and 55go-OH. The accessibility of the former molecule basal plane from either one or two sides results in different distortion of the carbon skeleton suggesting a new parameter affecting spectral properties of the molecules additionally to the effect of different oxidants. The side dependence is due to that the GOs carbon skeletons are formed by *sp³* configured atoms involved in the cyclohexanoid units of different isomorphic structures, which, in its turn, is responsible for peculiar bending of the pristine molecule basal plane presented in Fig. 1d [32, 33]. In contrast, skeletons of the framed molecules consist of *sp²* configured carbon atoms that still present compositions of benzenoid units, although with stretched C=C bonds.

## 3. Computational Details

The model structures in Fig. 1 were obtained by using the AM1 version of semi-empirical unrestricted Hartree-Fock (UHF) approach. The approach, applied to closed-shell molecules, provides obtaining solutions fully identical to the RHF one [34], which allows considering a complete set of the models at the same computational level. Optical absorption spectra were computed in one-geometry-point mode for the equilibrium structures presented in Fig. 1 in the framework of the restricted Hartree-Fock (RHF) semi-empirical approach, by adopting its ZINDO/S model implemented in Gaussian 09 package. Up to 80 excited states were included [35].

ZINDO/S as well as other RHF techniques are well tested and widely used to compute the optical properties of large molecules (see Ref. 36 and references therein). However, the techniques is strictly applicable to closed-shell molecules only while the consideration of excited states of open-shell molecules requires more sophisticated CI approaches, such as coupled cluster singles and doubles model (CC2) [37], EOMCCSD and CR-EOMCCSD(T) [38], unrestricted algebraic diagrammatic construction (UADC) scheme of second order [39, 40], UHF and UDFT calculations within the framework of the fragment molecular orbital (FMO) methods: (FMO-UHF) [41] and FMO-UDFT [42], respectively, as well as modified TD-DFT approaches [37, 38, 42, 43]. Hartree-Fock based techniques are much preferable due to substantial errors from TD-DFT calculations of excited states of $sp^2$ carbon system [37]. All the mentioned CI techniques [37, 38] are time-consuming and hardly applicable to large systems. More promising are UADC and FMO-UHF approaches [41-43] but even these techniques are not efficient enough to perform an extended computational experiment similar to suggested in the current paper for two sets of rGO and GO molecules. Leaving the correct calculations of open-shell rGO molecules for the future, we chose UHF-ZINDO/S approach combinations to perform a wished computational experiment. A large pool of available experimental data on GQDs optical spectra provides a reliable comparison with computational results so it becomes possible to make conclusion about the reliability of ZINDO/S calculations related to closed-shell molecules and to reveal the method drawbacks with respect to the open-shell ones. The latter is of particular interest since the computational technique is widely used involving open-shell molecules as well.

## 4. Results

*HF peculiarities of the molecules ground state*. The molecules shown in Figs. 1c and 1d are obtained in the course of stepwise hydrogenation [32] and oxidation [33] of pristine (5, 5) NGr molecule. Quantum chemical data of the studied molecules, relating to the ground state but having a direct link with the excited ones, are listed in Table 1. As seen in the table, all the framed rGO molecules are characterized by nonzero $N_D$ due to which they should be attributed to open-shell ones while the 'bulk' GO molecules are evidently closed-shell ones for which effectively unpaired electrons are absent. The rGO molecules additionally involve the average number of the unpaired electrons η related to one skeleton carbon atom. Both $N_D$ and η present a measure of the correlation of $p_z$ odd electrons and evidence their open-shell character. The energetic characteristics of the studied models involve the gap values obtained in the course of the total optimization under UHF algorithm $G_{UHF}$ and $G_{RHF}$ calculated under RHF algorithm for equilibrium UHF structures. The relative gap deviation $\delta G = (G_{UHF} - G_{RHF})/G_{UHF}$ marks changes in the HOMO-LUMO gap when going from UHF to RHF calculations at fixed molecular structure. As seen in Table 1, a drastic underestimation of the HOMO-LUMO gap $G_{RHF}$ is observed when molecules become open-shell. The feature belongs to one of the common peculiarities of the optical spectra calculations exhibiting a large red shift if open-shell molecules are considered in the framework of restricted versions of either HF or DFT algorithms [37, 38, 42].

Table 1. Total and average number of effectively unpaired electrons and HOMO-LUMO energy gap

| No | Molecules | $N_D$, e | $\eta$, e | HOMO-LUMO energy gap | | |
|---|---|---|---|---|---|---|
| | | | | $G_{UHF}$, eV | $G_{RHF}$ [1]), eV | $\delta G$, % |
| 1 | (5, 5)NGr (gr55) $C_{66}$ | 31.04 | 0.47 | 5.57 | 4.15 | 25.6 |
| 2 | 55gr-2H $C_{66}H_{44}$ | 12.23 | 0.18 | 6.35 | 3.10 | 51.2 |
| | 55gr-H $C_{66}H_{22}$ | 15.75 | 0.24 | 6.02 | 1.92 | 68.1 |
| | 55gr-O $C_{66}O_{22}$ | 16.02 | 0.24 | 5.81 | 2.94 | 49.4 |
| | 55gr-OH $C_{66}(OH)_{22}$ | 17.76 | 0.27 | 7.05 | 2.02 | 71.3 |
| 3 | go55-2side-OH | 0 | 0 | 8.43 | 8.43 | 0 |
| | go55-1side=O | 0 | 0 | 8.30 | 8.30 | 0 |
| | go55-2side=O | 0 | 0 | 8.26 | 8.26 | 0 |

[1]) One-point geometry calculations using UHF equilibrium structures.

*ZINDO/S excited states of closed-shell molecules.* In connection with the ground-state results discussed above, evidently, the results of the ZINDO/S calculated optical absorption spectra for open-shell and closed-shell molecules should be considered separately. We start from the results related to closed-shell $sp^3$ molecules of GOs belonging to models shown in Fig. 1d. The correctness of the approach application, characterized by $\delta G=0$, is doubtless so that the matter is about how well the computation results fit experimental evidences of optical spectra. Figure 2a presents a collection of calculated absorption spectra. As seen in the figure, the spectra are formed by optical excitation of different excited states presented by different $\delta$-bars and characterized by the location of the excitation on different atomic groups. Multiple different internal and external factors, such as various structural and dynamic inhomogeneities, result in the broadening of the spectral shape. This is usually taken into account via convolution of the $\delta$-spectra with Lorentzians of different full-width-half-maximum (FWHM) parameters. In the current study a standard convolution parameter of 0.1 eV was used.

Comparing spectra given in Fig. 2a, one can conclude that in all the cases they are located at high energy above 4 eV. This should be evidently expected due to $sp^3$ character of the molecule's carbon skeletons, which, as in the other numerous cases related to valence-saturated carbonaceous hydrooxides, provides absorption spectra located in the UV region [44]. The spectra exhibit a noticeable dependence on oxygen containing groups (OCGs) when substituting atomic oxygen, incorporated into carbonyls and epoxides in the molecule circumference area and on basal plane of 55gr=O species, respectively, by hydroxyls in the case of 55gr-OH. Less but still significant changes are observed when oxidation concerns either one or two sides of the skeleton. Altogether, the picture presented in the figure allows predicting a considerable inhomogeneous background of the empirical spectra of GOs caused by variation of OCGs exaggerated by one-side or two-side adsorption. This conclusion well correlates with experimentally observed absorption spectrum presented in Fig. 2b for the GO aqueous dispersion at room temperature. Evidently, the main part of empirical spectrum is located in the UV region while tremendously broadened. Actually, the temperature effect constitutes a large part of the broadening while the remaining is still large thus providing the characteristic dependence of the GO fluorescence on the excitation wavelength due to selective excitation of different emitting centers inhomogeneously distributed over the sample [23, 24]. This inhomogeneity was clearly observed in the inelastic neutron scattering spectra from different GO samples as well [13]. As for fluorescence spectra, once shifted at different excitations, they are located in visible region in the case of both aqueous dispersions [18] (Fig. 2b) and solid films [24] (Fig. 2c). As seen in Fig.2, significant Stock's shift between the lowest bound of the excited states energy and fluorescence spectra position is observed.

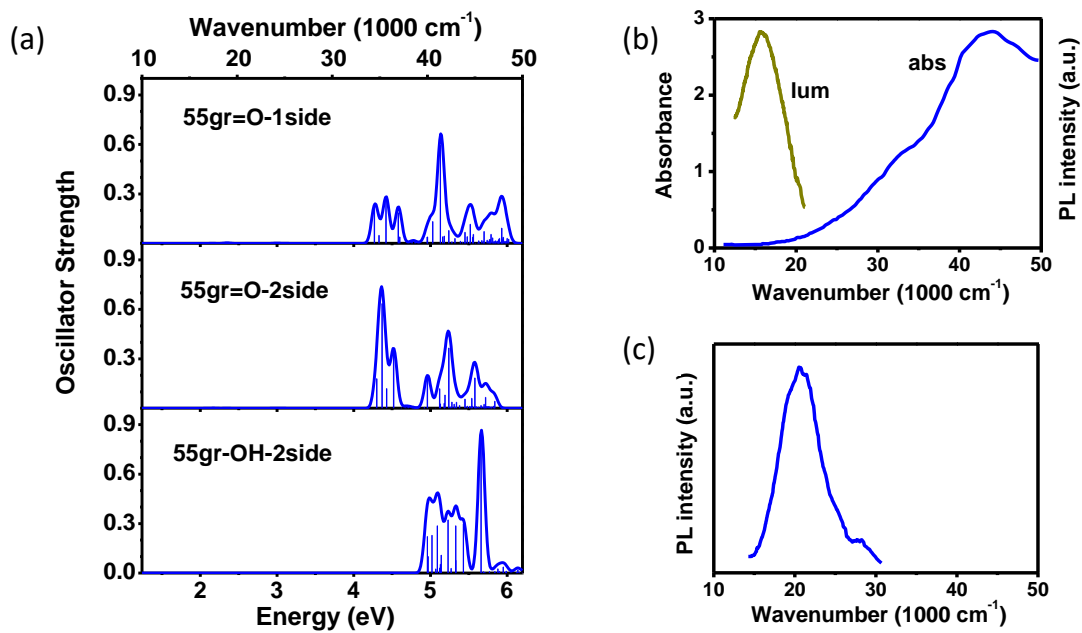

**Figure 2**: (a) ZINDO/S optical absorption spectra of GO molecules 55go=O-1side, 55go=O-2side and 55go-OH-2side from top to bottom, respectively. Pristine δ-band spectra are convoluted with a Lorentzian of 0.1 eV FWHM. (b) Absorption and fluorescence spectra of synthetic GO in water at room temperature; adapted from Ref. [18]. (c) Fluorescence spectrum of GO thin film at room temperature; adapted from Ref. [23].

*ZINDO/S excited states of open-shell molecules.* Figure 3a presents a collection of absorption spectra of four model rGO molecules shown in Fig. 1c. As in the previous case, the spectra are formed by optical transitions to a set of excited states differing by the location of the photoexcitations on the molecule's atomic groups. The pristine δ-spectra are Lorenzian convoluted by using FWHM of 0.1 eV. The spectra of single-atom framed 55gr molecules (55gr-H, 55gr=O and 55gr-OH) may be characterized by three regions of absorption, namely, the near-IR, visible and near-UV bands. The lowest-energy excitation is located at 0.4 eV (5000 cm$^{-1}$). Changing framing chemicals modifies the absorption slightly, concerning only not drastically important details. The largest change accompanies addition one more hydrogen atom to all edge carbon atoms of the circumference (see spectrum of 55gr-2H). Presented in the right column panels of Fig.3 are fluorescence spectra of rGO-Sh in crystalline toluene matrix at T=80K (Fig. 3b) and synthetic rGOs dispersed in different solution at room temperature (Fig. 3c). Additionally, the rGO-Sh absorption spectrum in CCl$_4$ dispersion at room temperature is shown in Fig. 3b.

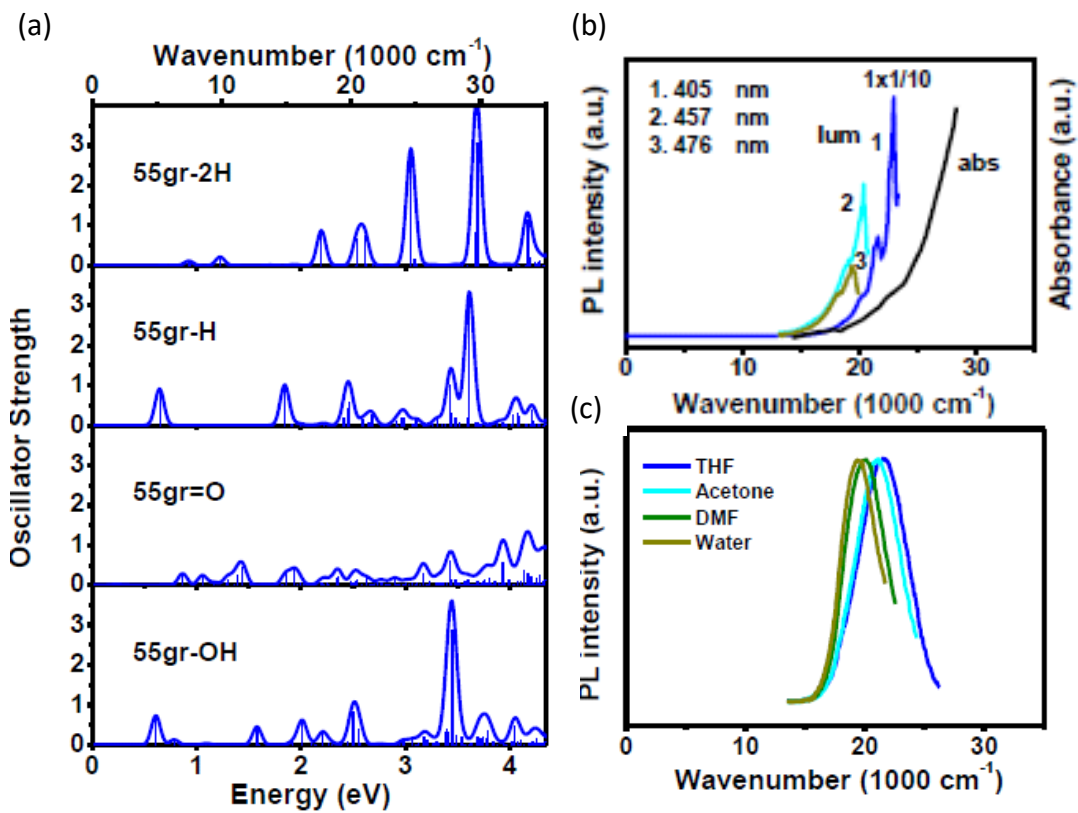

**Figure 3**: (a) ZINDO/S optical absorption spectra of rGO molecules 55gr-2H, 55gr-H, 55gr=O, and 55gr-OH from top to bottom, respectively. Pristine δ-band spectra are convoluted with a Lorentzian of 0.1 eV FWHM. (b) Absorption and fluorescence spectra of rGO-Sh in carbon tetrachloride dispersion at room temperature and crystalline toluene matrix at 80K, respectively; adapted from Ref. [26]. (c) Fluorescence spectrum of synthetic rGO in different solutions (see marking in the figure) at room temperature; adapted from Ref. [24].

As was discussed in Section 2, 55gh-H is a favorite for comparison with experimental spectra. Its chemical composition, described by formula $C_6O_{0.1}H_{1.6-0.7}$, shows that the specie is not pure graphene hydride but belongs to oxihydrides. Relating to 55gr-based molecules, its atomic content is better reproduced by formula $C_{66}O_1H_{21}$, which means the emergence of one oxygen atom among remaining 22 hydrogens. However, as seen in Fig. 3a, even complete substitution of hydrogen atoms by oxygen ones in 55gr=O does not cause too serious changes in the absorption spectra due to which the former still may keep its priority when comparing computational and experimental spectra. However, such a comparison of 55gr-H spectrum with that presented in Fig. 3b evidently reveals a complete disagreement since the latter has much in common with the experimental absorption spectrum of GO presented in Fig. 2b. Accordingly, the rGO spectrum is mainly located in the UV region with a low-energy bound on the boundary between the visible and UV region just providing the observation of the fluorescence spectrum in the visible region at a reasonable Stocks shift, as it actually takes place (see Figs. 3b and 3c) similarly to the situation concerning GO spectra. The observed discrepancy has a serious reason which will be considered in the next section.

## 5. Discussion

The main goal of the current section is to discuss the interrelation of calculated results with the empirical reality. As previously, the discussion concerning GOs and rGO will be performed separately. Addressing GOs, we are facing the following problems. The first is that the term graphene oxide covers an extremely large class of chemical products differing by chemical composition, size and shape. The last three factors significantly influence both absorption and fluorescence spectra making them variable in a large spectral region due to which the GO spectroscopy is spectroscopy of structurally inhomogeneous samples characterized by remarkable broadening. The next complication arises from studying the GO spectra in solutions (mainly in water) at room temperature. These two factors significantly exaggerate broadening, complementing it by red-shifting. Therefore, no standard empirical GO spectrum can be suggested to provide its correct comparison with a calculated one attributed to a fixed molecular structure at a quantitative level. We may operate with some average image of experimental reality only and look about the support of the general tendencies. From this viewpoint, the situation with GO spectroscopy is quite positive. According to experimental evidences [19, 46, 47], the GO absorption spectra is located in the UV-vis region and consists of an intense maximum at 250-400 *nm* (4.96-3.10 *eV*) followed with long wave tail up to near IR (Fig. 2b). Evidently, the maximum should be attributed to intense excitations located above 4 *eV* in the calculated GO absorption spectra in Fig. 2a. As for the long wave tail, a lot of experimentally provided complication may be suggested to explain its existence. Important factor is that there is no strong absorption in the visible region of empirical GOs that is in consent with calculated data. Similarly to absorption, the fluorescence spectrum is quite variable as well, nevertheless, presented by a single broad band. The band maximum position varies from 365 *nm* to 605 *nm* (3.40-2.05 *eV*) [19, 46, 47] (see Figs. 2b and 2c), which is strictly connected with the object under study. We will not speculate about the reasons of the variation since it requires a scrupulous analysis of all the stages of the experiments performed but draw attention on the fact that evidently GOs fluoresce in vis-UV region. The fact is well consistent with calculated spectra in Fig. 2a, according to which the emission should occur below 1.95-2.05 *eV* (640-610 *nm*) in the case of 55go=O-2side and 4.96 *eV* (250 *nm*) for 55go-OH-2side. A variable composition of the main oxidants, involving O, OH and COOH units, alongside with different size of GO molecules is possible to explain the observed variability of the experimental spectra. Thus, hydroxyls promote evidently more blue-shifted fluorescence than epoxy and carbonyl units.

The empirical situation concerning optical spectra of GQDs is very rich and diverse (see reviews [2, 22, 27]. It is caused by a large variety of rGO species (see, to name but a few [2, 8, 19, 48]), on the one hand, and a pronounced diversity in experimental techniques and conditions (solvents, temperature, various external actions), on the other. Among this variety there was a place for an undeniable uniqueness – the presence of the rGO-Sh spectrum that to a great extent can be attributed to a fixed molecule structure and thus considered as a standard spectrum. The uniqueness concerns fine structure fluorescence spectra of individual rGO-Sh molecules, fixed at low temperature in toluene matrix, whose size and chemical composition is well described by the 55gr-H model discussed above [26-28]. This fluorescence spectrum is presented in Fig. 3b at different excitation lengths. The spectrum collection reveals the possibility of selective excitation of different rGO molecules. Conservation of the spectrum shape followed by red-shifting at increased excitation wavelength evidences a predominant influence of the molecule size.

As seen in the figure, the fluorescence is observed in the visible region, which drastically contradicts the computational absorption spectrum given in Fig. 12.9a. Apparently, a considerable and unreliable red shift of the calculated 55gr-H absorption spectrum should be attributed to well known inability of RHF-based ZINDO/S calculations to correctly reproduce excited states of open-shell molecules [49]. Evidently, the disposition of absorption spectra in IR means a narrow HOMO-LUMO gap, which means its underestimation by the RHF formalism in the case of open-shell molecules. The data listed in Table 12.1 are coherent with the explanation. Actually, $G_{UHF}$ of 55gr-H is three times bigger than $G_{RHF}$ for the same molecular structure. Practically, the same ~three-fold changes are observed for all the rGO molecules. The difference evidently results in a mandatory blue shift of real absorption spectra with respect to the ZINDO/S ones due to which the fluorescence of the molecules should be observed in the visible region. At the same time, a gradual lowering of $G_{UHF}$ when going from 55gr-H to 1111gr-H and 1512gr-H based on (11, 11) NGr and (15, 12) NGr molecules [50], is well consistent with red shift of the fluorescence spectrum when the molecule size increases. In contrast, the dependence of excitation spectra on the chemical composition of the circumference framing of the same carbon skeleton of 55gr rGOs does not reveal any straight tendency. In this connection, we cannot ignore a special role of the states of edge atoms widely discussed in the literature (see Ref. 51 and references therein). In fact, the practice shows that their role is greatly exaggerated. Thus, in a series covering 55gr, 55gr-H and 55gr-2H molecules spin density on edge atoms changes drastically from 1.3-1.0 e in the parent (5,5) NGr to 0.42-0.22 e in 55gr-H and to zero in 55gr-2H due to complete saturation of the edge atom dangling bonds in the latter case. At the same time, $G_{UHF}$ takes the values 5.57 eV, 6.02 eV, and 6.08 eV, respectively, not revealing the presence of any drastic effect. Another case, $G_{UHF}$ changes from 6.02 eV to 5.81 eV when going from partially unsaturated to completely saturated dangling bonds in 55gr-H and 55gr=O, respectively, and then again reaches 7.05 eV in 55gr-OH with a partial insaturation. It would seem that once important, saturated/unsaturated states of the edge atoms should strongly affect the electronic states of the molecules while in practice the changes are not tendentious but a strictly individual. One can confidently assume that extension of the current computational experiment by using either algebraic-diagrammatic construction scheme (ADC) [41] or UHF fragment molecular orbital method (UHF-FMO) [42, 43], specially elaborated for the consideration of optical spectra of open-shell molecules, will allow clarifying the risen issues making the optical spectroscopy of open-shell molecules more transparent.

Evidently, the tree-fold lowering of $G_{UHF}$ under transformation to $G_{RHF}$ cannot be a universal law and may be a consequence of some other common properties of the studied molecules. Such an obvious common characteristic is a rectangular structure of the molecules of group 2 with clearly formed armchair and zigzag edges. If the structural architecture is changed, so that the edges are not clearly determined as it is in the case of large PAHs such as $C_{24}H_{12}$ and $C_{114}H_{30}$, the difference between $G_{UHF}$ and $G_{RHF}$ becomes less [50]. The finding is well consistent with a pronounced topochemistry peculiar to graphene structures in general [52], which allows suggesting a new topochemical effect of graphene molecules exhibited in the dependence of their optical spectra on the molecule shape.

## 6. Summary and Conclusions

In summary, we have presented results of an extended computational experiment covering quantum-chemical characteristics of the ground and excited states of a specially configured set of graphene-based molecules involving both closed-shell ('bulk' molecules) and open-shell (framed molecules) species. The choice of molecules was subordinated to meet the following requirements:
- The framed molecules belong to graphene oxihydrides with empty basal plane and are attributed to different rGOs that are the basic building elements of GQDs; they differ by chemical composition of the circumference area, size, and shape;
- Empirically determined structure and chemical composition of shungite GQDs lays the foundation of the deliberate choice of the selected framed molecules;
- The 'bulk' molecules are chosen among graphene oxides providing a direct link with the selected framed ones of the same carbon skeleton and circumference area;
- The molecules attribution to either open-shell or closed-shell type is performed on the basis of the UHF calculation of the total number of effectively unpaired electrons $N_D$.

Absorption spectra of the molecules were calculated in the framework of ZINDO/S approach applied to the UHF determined equilibrium molecular structures. Once restricted by ignoring electron correlation, the codes provide a reliable description of excited states of closed-shell molecules and once applied to the studied 'bulk' molecules, showed results well compatible with empirical observations concerning absorption and fluorescence spectra of graphene oxides. When applied to open-shell framed molecules, the tool revealed a drastic red shift of the absorption spectra and, as a result, a strong contradiction to the observation of GQDs fluorescence in visible region. The discrepancy was explained as a consequence of the electrons correlation ignorance, involved in the algorithm, followed by a severe underestimation of the HOMO-LUMO energy gap $G_{RHF}$ of open-shell molecules in contrast to $G_{UHF}$ predicted by UHF calculations. The latter values are well consistent with blue-green fluorescence of GQDs. Changing in the chemical composition, size and shape of the studied framed molecules does not affect the tendency of the $G_{RHF}$ considerable reduction in general while exhibiting a strict individuality in each case. In spite of the evident inadequacy concerning the ZINDO/S application to open-shell molecules, the obtained results are full of important information and are highly required for the costly and time-consuming computational experiments with the tools of higher CI level in use to be optimally designed.